\documentclass[reprint,aps,prl,  superscriptaddress,amsmath,amssymb,floatfix,letterpaper,longbibliography]{revtex4-1}
\usepackage{graphicx}
\usepackage{color}
\usepackage{times}

\usepackage[hypertexnames=false]{hyperref}
\hypersetup{colorlinks=true}
\usepackage[all]{hypcap} 

\usepackage{filecontents}

\newcommand{\IFPone}{IFP$_1$} 
\newcommand{\IFPtwo}{IFP$_2$} 

\begin{document}

\title{Rescuing a Quantum Phase Transition with Quantum Noise}

\author{Gu Zhang}
\affiliation{Department of Physics, Duke University, P.O. Box 90305, Durham, North Carolina 27708, USA}
\author{E. Novais}
\affiliation{Centro de Ci\'{i}ncias Naturais e Humanas, Universidade Federal do ABC,
Santo Andr\`{e}, SP 09210-580, Brazil}
\author{Harold U. Baranger}
\email{baranger@phy.duke.edu}
\affiliation{Department of Physics, Duke University, P.O. Box 90305, Durham, North Carolina 27708, USA}
\date{January 20, 2017}

\begin{abstract}
We show that placing a quantum system in contact with an environment can enhance non-Fermi-liquid correlations, rather than destroy quantum effects as is typical. The system consists of two quantum dots in series with two leads; the highly resistive leads couple charge flow through the dots to the electromagnetic environment, the source of quantum noise. While the charge transport inhibits a quantum phase transition, the quantum noise reduces charge transport and restores the transition. We find a non-Fermi-liquid intermediate fixed point for all strengths of the noise. For strong noise, it is similar to the intermediate fixed point of the two-impurity Kondo model. 
\end{abstract}

\maketitle
 
Quantum fluctuations and coherence are key distinguishing ingredients in quantum matter. Two phenomena to which they give rise, for instance, are \emph{quantum phase transitions} \cite{CarrBook,VojtaPhilMag06}, changes in the ground state of a system driven by its quantum fluctuations, and \emph{quantum noise} \cite{LeggettRMP87,WeissBook,IngoldNazarov92}, 
the effect on the system of quantum fluctuations in its environment, for no system is truly isolated. 
Understanding the intersection of these two topics---the effects of quantum noise on quantum phase transitions---is important for understanding quantum matter. It is natural to suppose that decoherence produced by the noise will suppress quantum effects, and in particular inhibit or destroy a quantum critical state. Indeed, a variety of calculations demonstrate this in both equilibrium \cite{LeggettRMP87,WeissBook,KapitulnikDissipQPTPRB01,LeHurPRL04,CazalillaDissipQPTLLPRL06,ChungQPTPRB07,HoyosVojtaPRL08,PollettiPRL12,CaiBarthelPRL13,*CaiBoseLiqPRL14} and non-equilibrium \cite{DallaTorrePRB12,*DallaTorreNP10,Foss-FeigNJP13,ChungNoneqDissipPRB13, Joshi1DxyPRA13,SiebererDynCritPRL13} contexts. 
There are also a few known cases that do not follow this rule \cite{WernerDissipIsingPRL05,NagyDomokosPRL15,MarinoDiehlPRL16}. Here, we present a striking counter-example to the notion that 
environmental noise necessarily harms quantum many-body effects: in the system we study, the addition of (equilibrium) quantum noise stabilizes a non-Fermi liquid quantum critical state. 

We discuss the phase diagram of two quantum dots connected to two leads in 
the presence of environmental quantum noise. The noiseless model has a quantum phase transition that is transformed into a crossover by charge transport across the double dot. We show that quantum fluctuations of the field associated with the source and drain voltage counteract this charge transport. 
The competition between these two processes 
restores the delicate balance of the quantum critical state. 
The result is that the quantum phase transition is rescued from the undesired crossover for \emph{any} strength of the noise. 

\begin{figure}[tb]
\centering \includegraphics[width=0.7\linewidth]{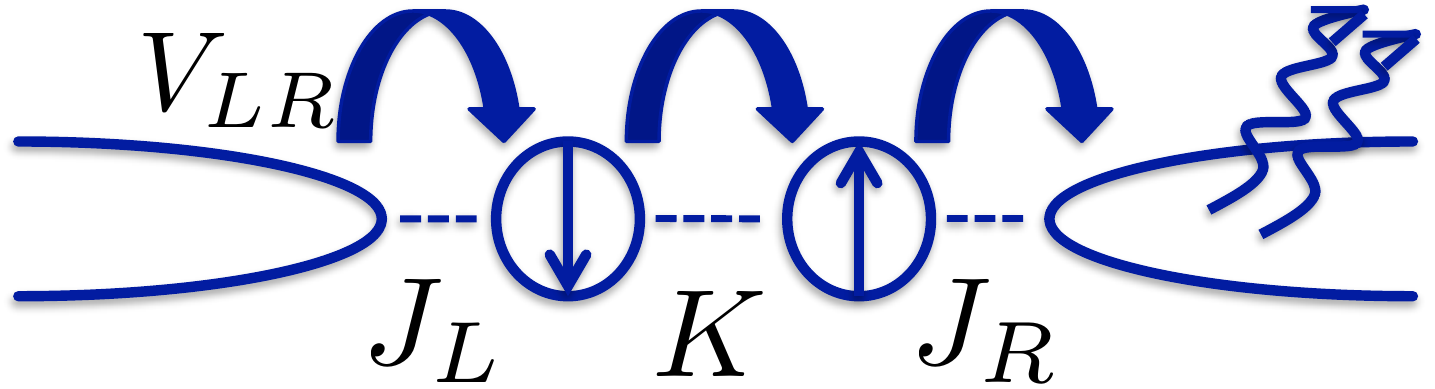} 
\caption{  Schematic of the system: two quantum dots coupled 
to left and right leads. $J_{L,R}$ and $K$ refer to the Kondo and exchange coupling strengths, respectively. $V_{LR}$ is the strength of direct charge transport between the leads. Dissipative modes in the leads are represented by wiggly arrows.}
\label{fig:system} 
\end{figure}

Our double quantum dot setup is shown schematically in 
Fig.\,\ref{fig:system}: two small dots are in series between two leads, 
labeled $L$ (left) or $R$ (right). The leads are resistive, thereby 
coupling the electrons to an ohmic electromagnetic environment. 
Experimentally, small double dots have been studied in several materials 
\cite{JeongChangScience01,ChorleyBuitelaarPRL12,KellerDGGNatP14,Bork2IKexptNP11,Spinelli2IKexptNC15}, and the 
effect of the environment on transport in simpler systems has been 
recently studied in detail \cite{ParmentierPierre11,Mebrahtu12,Mebrahtu13,JezouinPierre13}, 
including transport through a single quantum dot \cite{Mebrahtu12,Mebrahtu13}.
Thus, all the necessary 
ingredients for an experimental study of our system are available.

\emph{Model for dots and leads---}The model has three parts: leads, dots, 
and electromagnetic environment. Following 
standard procedures, we linearize the spectrum of each lead, notice that a one-dimensional subset of electrons couples to each dot, and represent it using chiral fermions by analytic continuation with open boundary conditions  
\cite{[{}][{ pp. 351-368.}]GogolinBook}. The resulting lead Hamiltonian 
is the sum of four free Dirac fermions, 
\begin{equation}
H_\text{leads}^{0}= \sum_{\alpha,\sigma} \int_{-\infty}^{\infty}\!\!dx\,\psi_{\alpha,\sigma}^{\dagger}(x)i\partial_{x}\psi_{\alpha,\sigma}(x),\label{eq:Hleads}
\end{equation}
where $\alpha$ and $\sigma$ are the lead and spin labels and 
both the Fermi velocity and $\hbar$ are set to unity. 

For the dots, we consider the Coulomb blockade regime in which charge fluctuations are suppressed and the electron number is odd \cite{NazarovBook}. 
The single-level Anderson model is suitable for each dot, as the spacing between levels in the carbon nanotube dots is large \cite{Mebrahtu12,Mebrahtu13}. 
Each dot, then, has a low energy spin-$\frac{1}{2}$ degree of freedom, $\vec{S}_{\alpha}$.
Projecting onto this low-energy subspace via a second-order Schrieffer-Wolff transformation produces two Kondo-like terms with couplings $J_{L,R}$ and a spin-spin anti-ferromagnetic interaction with coupling $K$: 
\begin{equation}
H_\text{dots}=
J_{L}\vec{s}_{L}(0)\cdot\vec{S}_{L}+J_{R}\vec{s}_{R}(0)\cdot\vec{S}_{R}
+ K\vec{S}_{L}\cdot\vec{S}_{R},\label{eq:Hdots}
\end{equation}
where 
$\vec{s}_{\alpha}\!=\!\psi_{\alpha}^{\dagger}(0)\vec{\sigma}\,\psi_{\alpha}(0)$ is the spin-density in the lead at the point connected to the dot. Though none of our results depend on left-right symmetry, we take $J_L\!=\!J_R$ for simplicity. 

Charge transfer between the two leads is key to the physics of this system \cite{JonesKotliarMillisPRB89,GeorgesMeirPRL99,ZarandDotsPRL06,SelaPairTunnelPRL09a, MaleckiDoubdotPRB10,Logan2CKPRB11}. 
The effective hopping between the leads that arises from the third-order Schrieffer-Wolff transformation of 
the original Anderson model must be added \cite{Logan2CKPRB11}:
\begin{eqnarray}
H_{LR} & = & V_{LR} \left[
\big(\psi_{L\uparrow}^{\dagger}\psi_{R\uparrow}^{\,} 
+ \psi_{R\downarrow}^{\dagger}\psi_{L\downarrow}^{\,} \big)
S_{L}^{-} S_{R}^{+} \right. \nonumber \\
 & + & \big( \psi_{L\uparrow}^{\dagger}\psi_{R\uparrow}^{\,} 
 + \psi_{L\downarrow}^{\dagger}\psi_{R\downarrow}^{\,} \big)
 S_{L}^{z}S_{R}^{z} \label{eq:h3}\\
 & + & \left. \big(\psi_{L\uparrow}^{\dagger}\psi_{R\downarrow}^{\,}
 - \psi_{R\uparrow}^{\dagger}\psi_{L\downarrow}^{\,} \big)
 \left(S_{L}^{z}S_{R}^{-}-S_{L}^{-}S_{R}^{z}\right)\right] 
 + \mbox{h.c.}, \nonumber 
\end{eqnarray}
where $x\!=\!0$ for the lead operators \cite{SupMat}. 
\nocite{HewsonBook,WongAffleck94,OshikawaAffleck97}
This 
form is obtained because moving an electron across the dots necessarily involves the dot spins. 
Much of the physics added by (\ref{eq:h3}) is obtained from a simpler direct hopping, 
$\hat{H}_{LR}=\hat{V}_{LR}\psi_{L\sigma}^{\dagger}(0)\psi_{R\sigma}^{\,}(0)+\mbox{h.c.}$  \cite{Logan2CKPRB11,MaleckiErratumPRB11}.
We therefore simplify the discussion by using $\hat{H}_{LR}$ rather than $H_{LR}$ when possible \cite{SupMat}. 

The final ingredient in our system is the ``quantum noise.'' Quantum fluctuations of the source and drain voltage 
require a quantum description of the tunneling junction \cite{IngoldNazarov92,NazarovBook}.
The standard procedure is to introduce junction charge and phase fluctuation
operators that are conjugate to each other and (bilinearly) coupled to modes of the ohmic environment with resistance $R$. Treating the latter as a collection of harmonic oscillators with the desired impedance, we write the environment as a free bosonic field, $H_{\varphi}^{0}\!=\!\int\frac{dx}{4\pi}(\partial_{x}\varphi)^{2}$, which is excited in a tunneling event through the charge-shift operator $e^{i\sqrt{2r}\varphi(0)}$ \cite{IngoldNazarov92}. 
Such a shift operator is added to every term in $H_{LR}$ according to
\begin{equation}
\psi_{L\sigma}^{\dagger}\psi_{R \sigma}^{\,} \to 
e^{i\sqrt{2r}\varphi(0)} \psi_{L\sigma}^{\dagger}\psi_{R\sigma}^{\,},
\label{eq:add_dissipation}
\end{equation}
where $r\!=\!R e^2/h$ is the dimensionless resistance. 
The environment does not modify the second-order exchange couplings Eq.\,(\ref{eq:Hdots}) because those virtual processes occur on the very short time scale of the inverse charging energy \cite{FlorensPRB07}, typically smaller than the time scale of the environment.
This model of noisy tunneling has been used previously in work on a resonant level \cite{ImamAverinPRB94,LeHurLiPRB05}, including in our own work \cite{Mebrahtu12,Mebrahtu13,LiuRLdissipPRB14,Huaixiu14_1-G}, and for a quantum dot in the Kondo regime \cite{FlorensPRB07}.
In summary, the starting point of our discussion is the Hamiltonian
\begin{equation}
H=H_\text{leads}^{0}+H_{\varphi}^{0}+H_\text{dots}+{H}_{LR}\left(r\right).\label{eq:full_hamiltonian}
\end{equation}

\emph{Quantum phase transition or crossover?}---First, we bosonize the chiral fermions describing the leads, Eq.\,(\ref{eq:Hleads}), thereby introducing chiral bosonic fields $\phi_{\alpha,\sigma}$ \cite{GogolinBook,MaleckiDoubdotPRB10,SupMat}. 
One can then see that the ultraviolet fixed point, described by 
$H_\text{leads}^{0}+H_{\varphi}^{0}$,
is unstable. There are two important energy scales 
connected to this instability: the Kondo temperature, $T_K$, associated with the screening of each dot by its own lead, and
the ``crossover temperature,'' $T^*\!<\!T_K$  \cite{MaleckiDoubdotPRB10,MaleckiErratumPRB11}.

\begin{figure}
\includegraphics[width=0.95\columnwidth]{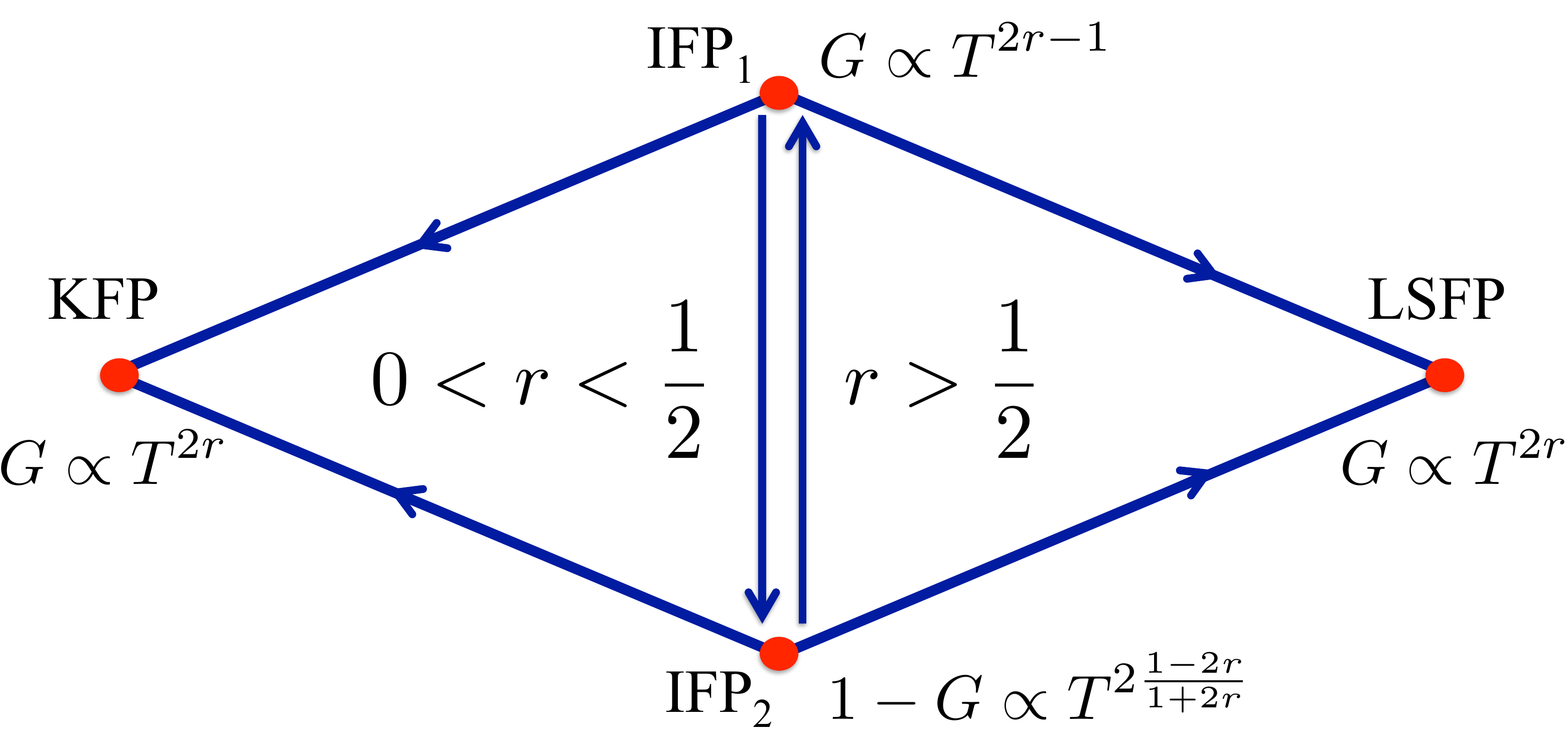}
\caption{ 
Stability diagram for different noise strengths $r$, and the 
temperature dependence of the conductance at each fixed point. The KFP and 
LSFP are stable for any non-zero $r$, while the nature of the IFP changes as 
a function of $r$. For $r\!<\!1/2$, the  intermediate state is controlled by 
\IFPtwo---the fixed point that evolves from one of the $r\!=\!0$ Fermi-liquid 
fixed points. In contrast, for $r\!>\!1/2$, \IFPone---which evolves from the two-impurity-Kondo IFP---is relevant. For $r\!=\!1/2$ a line of fixed points connects 
\IFPone\ to \IFPtwo. } 
\label{fig:flow-dissip}
\end{figure}

To explain $T^*$, we start by considering $V_{LR}\!=\!0$, yielding the two-impurity Kondo model. 
For $T\!<\!T_K$, there are two Fermi-liquid phases with a 
critical coupling that separates them \cite{JonesKotliarMillisPRB89,Affleck2IKPRB95}, denoted by 
$K_c$. 
(i)~For $K\!>\!K_c$, the two dots 
become maximally entangled in a singlet state---the \emph{local-singlet 
phase} controlled by a fixed point, denoted 
LSFP, with a scattering phase shift of $0$. 
(ii)~For $K\!<\!K_c$, each dot becomes maximally entangled with its respective lead, forming two decoupled Kondo singlets---the 
\emph{Kondo phase} controlled by a fixed point, denoted KFP, phase shift of $\pi/2$. The fact that 
the phase shifts are different implies the existence of an 
intermediate (unstable) fixed point \cite{Affleck2IKPRB95,Ludwig97p565}, 
which we call \IFPone\ (see Fig.\,\ref{fig:flow-dissip}).

Inter-lead tunneling, $V_{LR}\!\neq\!0$, changes the behavior dramatically. 
In the absence of dissipation, $r\!=\!0$, it is known that $H_{LR}$ destabilizes \IFPone\ \cite{SelaExactTransPRL09,ZarandDotsPRL06,SelaPairTunnelPRL09a, MaleckiDoubdotPRB10,Logan2CKPRB11}, becoming effective below a scale $T^*$. 
The low-energy physics is described by Fermi-liquid Hamiltonians with scattering phase shift varying from $\delta\!=\!0$ to $\pi/2$ depending on the initial values of the couplings \cite{GeorgesMeirPRL99,SelaPairTunnelPRL09a,MaleckiDoubdotPRB10,MaleckiErratumPRB11}. 
The finite temperature conductance is 
$G=G_0\sin(2\delta)[1\!-\!\kappa(T/T^*)^{2}]$,
where $G_{0}\!=\!2e^{2}/h$ and $\kappa$ is non-universal.
Therefore, for $T\!<\!T^*$, 
the quantum phase transition of the two-impurity Kondo model is 
transformed into a crossover between the Kondo and local-singlet regimes.



\emph{Quantum noise effects}---Close to the KFP and LSFP, the tunneling Hamiltonian in the absence of noise, 
$H_{LR}$,
is a marginal operator \cite{SelaPairTunnelPRL09a,MaleckiDoubdotPRB10,MaleckiErratumPRB11}: without noise, 
any bilinear operator that transfers charge between the 
leads is marginal at these two fixed points. A key effect of the noise is 
that 
the scaling dimension of such an operator 
increases, making it irrelevant. In the tunneling operator 
$H_{LR}$, the increase is caused by the exponential 
charge-shift operator introduced in Eq.\,(\ref{eq:add_dissipation}).  The line of Fermi-liquid fixed points existing at 
$r\!=\!0$ is then destroyed. The 
conductance around the KFP and LSFP follows from 
perturbation theory in the tunneling, leading to 
$G(T)\sim T^{2r}$ 
\cite{IngoldNazarov92,KaneFisherPRL92,*KaneFisherPRB92}. 

The new found stability of the Kondo and local-singlet fixed points
with respect to tunneling demands once again the existence
of an intermediate fixed point. We denote this ``dissipative intermediate fixed point'' by \IFPtwo, as shown in Fig.\,\ref{fig:flow-dissip}.

\IFPtwo\ occurs for the same value of $K$ as \IFPone, namely $K\!=\!K_c$, as we now show. It is known that the effect of a resistive environment on a bilinear tunneling operator is connected to the partition noise produced by the tunneling \cite{LevyYeyatiPRL01,SafiSaleurPRL04}: when there is no partition noise, the environment is not excited by the current and so has no effect. This is the case at $K\!=\!K_c$: the phase shift is $\delta\!=\!\pi/4$, so the zero temperature conductance is $G\!=\!2e^2/h$ and the transmission is unity. Thus, there is no partition noise: from the line of $r\!=\!0$ Fermi-liquid fixed points, this fixed point survives at non-zero $r$ and is, in fact, \IFPtwo. We now turn to characterizing both IFP's in detail.

\emph{Effective Hamiltonian at the intermediate fixed points}---In order to
derive an effective Hamiltonian at the critical coupling, $K\!=\!K_c$, we follow the dissipationless discussion of J.\,Gan in Ref.\ \cite{GanPRB95}. First, we define new bosonic fields,  
\begin{eqnarray}
\label{bosonic_fields}
\phi_{c/s} & = & \left(\phi_{L\uparrow} \pm \phi_{L\downarrow} + \phi_{R\uparrow} \pm \phi_{R\downarrow}\right)/2, \nonumber \\
\phi_{cf/sf} & = & \left(\phi_{L\uparrow} \pm \phi_{L\downarrow}-\phi_{R\uparrow} \mp \phi_{R\downarrow}\right)/2.
\end{eqnarray}
Physically, $\phi_c$ ($\phi_s$) represents the total charge (spin) in the leads, and $\phi_{cf}$ ($\phi_{sf}$) represents the corresponding difference between the left and right leads.
Next, one applies the unitary rotation $U\!=\!e^{-i\left(S_{1}^{z}+S_{2}^{z}\right)\phi_{s}(0)}$, thus dressing the spin states and making the exchange couplings anisotropic \cite{SupMat}. 
A key aspect of the physics at the IFP's is the degeneracy in the dots between the two dressed spin states 
$|0\rangle\equiv(|\!\!\uparrow\uparrow\rangle\!+\!|\!\!\downarrow\downarrow\rangle)/\sqrt{2}$ and 
$|1\rangle\!\equiv\!(|\!\!\uparrow\downarrow\rangle\!-\!|\!\!\downarrow\uparrow\rangle)/\sqrt{2}$ 
that leads to an effective Kondo problem with Kondo temperature $\tilde{T}_K$ \cite{GanPRB95}. It is convenient to introduce Majorana operators 
$a$ and $b$ 
for this two-dimensional Hilbert space (the string from the Jordan-Wigner transformation is incorporated into the lead operators). 
The end result \cite{GanPRB95} is 
an effective Hamiltonian for $K\!=\!K_c$ 
\cite{SupMat},
\begin{eqnarray}
\lefteqn{H_{K=K_c} =\!\!\!\sum_{\beta=\left\{ c,s,sf,cf,\varphi\right\}}
\!\!\!\!\!\!H_{\beta}^{0}
+2i\tilde{J}\frac{F_{sf}}{\sqrt{\pi \alpha}}\sin\left[\phi_{sf}(0)\right]a} \hspace{0.6in}
\nonumber \\
 & + &  2\tilde{V}_{LR} \frac{F_{cf}}{\sqrt{\pi\alpha}}
\cos\left[\phi_{cf}(0)
 +\sqrt{2r}\varphi(0)\right]b, \quad\;
\label{eq:Hboson1}
\end{eqnarray}
where 
$F_{sf},F_{cf}$ are Klein factors, $\alpha$ is of order the inverse cutoff, 
$\tilde{J}$ is the renormalized Kondo coupling, and 
$\tilde{V}_{LR}$ is the renormalized charge tunneling strength.  
Explicit expressions for $\tilde{J}$ and $\tilde{V}_{LR}$ are given in the supplemental material \cite{SupMat}.
 
The bosonic fields can be further untangled by 
performing a rotation that combines 
the field representing charge transfer between the leads, $\phi_{cf}$, with the environmental noise, $\varphi$: 
$\tilde{\phi}_{cf}\!\equiv\!(\phi_{cf}+\sqrt{2r}\,\varphi)/\sqrt{1+2r}$ and 
$\tilde{\varphi}\!\equiv\!(\sqrt{2r}\,\phi_{cf}-\varphi)/\sqrt{1+2r}$. 
%
The symmetries of the model are explicitly shown by defining six
Majorana fermionic fields \cite{SelaPairTunnelPRL09a,MaleckiDoubdotPRB10,Ludwig97p565}
with Ramond boundary conditions,
$\chi_{\beta}^{1,2}(0^{+})=\chi_{\beta}^{1,2}(0^{-})$:
$\chi_{\beta=\left\{c,s,sf\right\}}^{(1)}(x)=\frac{F_{\beta}}{\sqrt{\pi\alpha}}\sin\left[\phi_{\beta}(x)\right]$ and 
$\chi_{\beta=\left\{c,s,sf\right\}}^{\left(2\right)}(x)=\frac{F_{\beta}}{\sqrt{\pi\alpha}}\cos\left[\phi_{\beta}(x)\right]$.
Because the boundary interaction $2i\tilde{J}\chi_{sf}^{(1)}(0)a$
has scaling dimension $1/2$ ($a$ is an impurity operator), 
$\tilde{J}$ flows to strong coupling \cite{SelaPairTunnelPRL09a,MaleckiDoubdotPRB10}.
$\chi_{sf}^{(1)}$ then incorporates $a$ and  can be expressed as a simple change of boundary condition from Ramond to Neveu-Schwarz: 
$\chi_{sf}^{(1)}\left(0^{+}\right)\!=\!- \chi_{sf}^{(1)}\left(0^{-}\right)$ \cite{Ludwig97p565}.

The effective IFP Hamiltonian can, thus, be written in terms of
six free Majorana fields---five with Ramond and one with Neveu-Schwarz boundary condition---one free bosonic field ($\tilde{\varphi}$), 
and a boundary sine-Gordon model for $\tilde{\phi}_{cf}$: 
\begin{eqnarray}
H_\text{IFP} & = & \sum_{j=1}^{5}\!\int\! \frac{dx}{2}
\chi_{j}(x) i\partial_{x} \chi_{j}(x) 
+ \!\int\! \frac{dx}{2}\chi_{sf}^{(1)}(x)i\partial_{x}\chi_{sf}^{(1)}(x)
\nonumber \\
 & + &  \int\frac{dx}{4\pi} [ \partial_{x}\tilde{\varphi}(x) ]^{2} 
       +\int\frac{dx}{4\pi}[ \partial_{x}\tilde{\phi}_{cf}(x) ]^{2}
\nonumber \\ & + & 
 2i\,\tilde{V}_{LR}\frac{F_{cf}}{\sqrt{\pi \alpha}}\cos\left[\sqrt{1+2r}\,\tilde{\phi}_{cf}(0)\right]b.\label{eq:nIFP}
\end{eqnarray}
This Hamiltonian has an inherent ${\it SO}(5)\!\times\!U(1)$ symmetry from the five Majorana fields and the dressed dissipation field $\tilde{\varphi}$.
With regard to the dot degrees of freedom, while Majorana mode $a$ is effectively incorporated into the leads, mode $b$ is coupled to the charge transport. For the two-impurity Kondo model, $\tilde{V}_{LR}\!=\!0$ and $b$ is a decoupled Majorana zero mode.


\emph{Dependence of IFP on dissipation}---The boundary sine-Gordon model, which is the last element in Eq.\,(\ref{eq:nIFP}), is well known to have a quantum phase transition \cite{GuineaPRB85,Affleck98p2761,GogolinBook} 
as the parameter in the boundary term varies, in our case $r$. The simplest description of this transition is via the scaling equation,
$\frac{d\tilde{V}_{LR}}{d\ell}=\Big(\frac{1}{2}-r\Big)\tilde{V}_{LR}$, 
which results from noticing that the scaling dimension of the operator 
$\cos[\sqrt{1+2r}\,\tilde{\phi}_{cf}(0)]$ is $(1+2r)/2$ \cite{GogolinBook}. 
There are three distinct scaling behaviors depending on the value of $r$. 

For \emph{weak} dissipation, $r\!<\!1/2$, $\tilde{V}_{LR}$ grows. As in the $r\!=\!0$ case \cite{SelaPairTunnelPRL09a,MaleckiDoubdotPRB10}, the cosine gets pinned at 
a particular value. The fixed point Hamiltonian is obtained by changing the 
boundary condition on $\tilde{\phi}_{cf}$ at $x\!=\!0$ from Dirichlet 
[for open boundary conditions on the fermionic fields in Eq.
\,(\ref{eq:Hleads})] to Neumann \cite{Affleck98p2761}. \IFPtwo\ is the 
corresponding fixed point; it develops from the $\delta\!=\!\pi/4$ Fermi-liquid 
fixed point \cite{SelaPairTunnelPRL09a} of the dissipationless case. 


The leading irrelevant operator at \IFPtwo\ is, because of the change in boundary condition, simply the dual of the relevant operator at \IFPone\ that causes $\tilde{V}_{LR}$ to grow \cite{GogolinBook, WeissBook}. Its scaling dimension is 
$2/(1+2r)$---the inverse of that of the cosine operator above.
The temperature dependence of the conductance is therefore expected to be
\cite{SupMat}   
\begin{equation}
G \sim G_0 \big[ 1-\gamma T^{2(1-2r)/(1+2r)} \big]
\qquad {\rm (at\ IFP_2)}
\end{equation}
with 
$\gamma$ a non-universal constant. We see that modification of the boundary interaction by dissipation introduces a Luttinger-liquid-like character. In 
addition to the conductance, the non-Fermi liquid nature of this fixed 
point is also manifest in its residual boundary entropy,
which can be shown to be $\ln g_{\text{IFP}_2}\!=\!\frac{1}{4}\ln\left(1+2r\right)$ \cite{SupMat}.

The break down of scaling 
(i.e.\ when $\tilde{V}_{LR}$ becomes of order one) defines the crossover temperature,
$T_\text{noise}^{LR}\!\approx\!T_{K}\;\big(\tilde{V}_{LR}^0 \big)^{2/(1-2r)}$,
in terms of the initial value of tunneling from left to right, $\tilde{V}_{LR}^0$ \cite{TKappears}. 
For higher temperatures, $T_\text{noise}^{LR}\!<\!T\!<\!T_{K}$, the physics is 
controlled by the $\tilde{V}_{LR}\!=\!0$ fixed point, \IFPone, as 
$\tilde{V}_{LR}$ is initially small. For lower temperatures, 
$T\!<\!T_\text{noise}^{LR}$, the 
physics is controlled by \IFPtwo. 

To study the effect of deviations of the antiferromagnetic coupling $K$ from $K_c$, we follow the discussion in Refs. \cite{MaleckiDoubdotPRB10,MaleckiErratumPRB11} and define the crossover 
temperature 
$T_{\delta K}\!=\!a(K-K_c)^{2}/T_K$,
where $a$ is a dimensionless constant. If 
$T_\text{noise}^{LR}\!<\!T_{\delta K}$, the low energy physics will be governed 
by the KFP or LSFP. However, for 
$T_{\delta K}\!<\!T_\text{noise}^{LR}$ an experiment would initially observe a 
rise in the conductance due to proximity to \IFPtwo\ before the crossover 
to the Kondo or local-singlet physics took over (for which $G\!\to\!0$). 
Using the remarkable tunability of quantum dots, access to the regime 
$T_{\delta K}\ll T_\text{noise}^{LR}$ is possible, in which case the power 
law approach of the conductance to the quantum limit $G_0$, given above, 
should be observable. Indeed, a strong-coupling fixed point with similar 
properties has recently been studied experimentally in a single dissipative 
quantum dot \cite{Mebrahtu12,Mebrahtu13}. 

In sharp contrast, for \emph{strong} dissipation, $r\!>\!1/2$, $\tilde{V}_{LR}$ shrinks, and the properties of the system are controlled by \IFPone. 
The boundary condition on the field  $\tilde{\phi}_{cf}$ remains the Dirichlet condition. The scaling dimension of the boundary sine-Gordon term implies that the conductance decreases at low temperature according to \cite{SupMat} 
\begin{equation}
G\sim T^{2r-1} \qquad {\rm (at\ IFP_1)}.
\end{equation}
The non-Fermi liquid nature of this fixed point is further shown by the residual boundary entropy, $\ln g_{\text{IFP}_1}\!=\!\frac{1}{4}\ln[4/(1+2r)]$ \cite{SupMat}, and by the decoupling of the $b$ Majorana in the dots as the last term in Eq.\,(\ref{eq:nIFP}) flows to zero.

\IFPone\ evolves from the intermediate fixed point of the two-impurity Kondo 
model ($r\!=\!0$). Formally, however, \IFPone\ is a distinct fixed point---the 
residual boundary entropy, for instance, depends on $r$. 
Nevertheless, for 
reasonable values of $r\!\sim\!1/2$, this system can emulate the 
physics of the two-impurity Kondo model: 
the ${\it SO}(7)$ symmetry manifest in the Majorana 
fields \cite{Ludwig97p565,MaleckiDoubdotPRB10}, for instance, 
is restored asymptotically. Any 
observable not directly related to charge transfer between the leads, such as the magnetic susceptibility, will have the same behavior in the two models. 

The crossover temperature to the KFP or LSFP, $T_{\delta K}$, is given 
by the same expression as in the weak noise case. Thus, for 
$T_{\delta K}\!<\!T\!<\!T_{K}$ the physics of \IFPone, bearing strong 
resemblance to that of the two-impurity Kondo model, will be experimentally 
accessible. 

Finally, the \emph{borderline} $r\!=\!1/2$ case is particularly interesting. 
The cosine in Eq.\,(\ref{eq:nIFP}) is exactly marginal \cite{Callan94p417}, 
corresponding to an $SU(2)$ chiral symmetry. Hence, we can replace the cosine by the Abelian chiral current $\partial_{x}\tilde{\phi}_{cf}$ \cite{Affleck98p2761}. The model becomes quadratic and the conductance can be calculated exactly \cite{KaneFisherPRL92,GogolinBook}---$G$ depends on the initial value $\tilde{V}_{LR}^0$ and so is not universal. 
The exactly marginal operator creates a line of fixed points connecting \IFPone\ to \IFPtwo, all with residual boundary entropy
$\frac{1}{4}\ln2$.
The line is unstable to deviations from the critical coupling $K_c$;
as in the previous cases, $T\!<\!T_{\delta K}$ leads to flow toward the KFP or LSFP. Even at $K_c$, corrections to the effective Hamiltonian (\ref{eq:nIFP}) will presumably cause flow away from this line at the lowest temperatures (which we have not analyzed); however, because their initial strength is very small, the cross-over temperature $T^*$ to see these effects will be very low. Thus, in a wide range of temperatures, $T^*\!<\!T\!<\!T_K$, the properties of the line of fixed points could be seen experimentally, varying $V_{LR}^0$ to move among them.

\emph{Conclusions}---We have presented an example in which the introduction of a quantum environment reveals a quantum phase transition previously hidden under a crossover: \emph{the quantum noise has rescued the quantum phase transition.} There are two quantum critical points (Fig.\,\ref{fig:flow-dissip}): one dominant for weak dissipation (\IFPtwo, $r\!<\!1/2$) and the other at strong dissipation (\IFPone, $r\!>\!1/2$)---this latter fixed point is similar to that of the two-impurity Kondo model. 


A broader view is obtained by connecting to the idea of ``quantum frustration of decoherence'' of a qubit \cite{CastroNetoQfrustPRL03,NovaisQfrustPRB05}: 
a quantum system acted upon by \emph{two} processes that are at cross purposes may retain more coherence than if acted upon by just one. 
The quantum system to be protected here is the non-Fermi-liquid quantum critical state delicately balanced between the KFP and LSFP, a striking signature of which is the decoupled, and so completely coherent, Majorana mode. 
Charge transfer between the electron reservoirs associated with the leads
is the first process acting on the system, one that completely destroys the delicate quantum state and the coherence of the Majorana mode.
Adding the quantum noise produced by the resistive EM environment impedes the deleterious effect of the first process, rendering the coherent Majorana zero mode again manifest at \IFPone. 
Thus, the quantum coherence of the delicate many-body state survives due to the ``quantum frustration'' of these two processes. 

This quantum critical state is highly non-trivial and clearly unstable toward the KFP and LSFP, but it has experimental consequences in a wide temperature range. We  emphasize that measurements of the conductance near \IFPone\ and \IFPtwo\ are experimentally feasible at this time---similar amounts of tuning have been used successfully, for instance, in recent experiments \cite{Mebrahtu12,Mebrahtu13}. An experimental study along these lines would directly contradict the general notion that more noise leads inevitably to less quantum many-body behavior. 

\begin{acknowledgments}
We thank T.\ Barthel, G.\ Finkelstein, S.\ Florens, and M.\ Vojta for helpful conversations. The work in Brazil was supported by FAPESP 
Grant 2014/26356-9. The work at Duke was supported by the U.S.\ DOE Office of Science, Division of Materials Sciences and Engineering, under Grant 
No.\ DE-SC0005237. 
\end{acknowledgments}

\widetext
\clearpage

\renewcommand{\bibnumfmt}[1]{[S#1]}
\renewcommand{\citenumfont}[1]{S#1}
\global\long\def\theequation{S\arabic{equation}}
\global\long\def\thefigure{S\arabic{figure}}
\setcounter{equation}{0}
\setcounter{figure}{0}

\begin{center}
\textbf{\large Supplemental Material for ``Rescuing a Quantum Phase Transition with Quantum Noise''\\
	\vspace{5pt}
Quantum Critical Point''}\\
\vspace{15pt}
Gu Zhang, E. Novais, and Harold U. Baranger\\
(Dated: October 28, 2016)
\end{center}
\vspace{10pt}

In this Supplemental Material, we show some details concerning four points: (i) coupling strength expressions when dot charge degrees of freedom are included, (ii) derivation of the effective Hamiltonian for $K\!=\!K_c$, Eq.\,(7) of the main text, (iii) conductance at the intermediate fixed points, and (iv) calculation of the boundary entropy at \IFPone\ and \IFPtwo.

\section{I.\hspace{10pt}Kondo, exchange, and charge transport coupling strengths from  model with dot charge degrees of freedom}

In the main text, we assume that an odd number of electrons occupies each quantum dot and project onto the low-energy spin subspace of the dot degrees of freedom. However, one can take a step back to a model that includes charge fluctuations to the extent that change in occupancy by $\pm1$ is included and then find the exchange and charge transport amplitudes in Eqs.\,(2) and (3) in terms of the extended model. Thus, we assume that both the charging energy and the mean level separation of the dots is large, as is the case in carbon nanotube quantum dots \cite{Mebrahtu12s,Mebrahtu13s}, and so model our system by a single level Anderson model:
\begin{equation}
\begin{aligned}
H=H^0_{\rm leads}+\epsilon \sum_{\alpha \sigma}  n_{\alpha \sigma}
+U \sum_{\alpha} n_{\alpha \uparrow} n_{\alpha \downarrow}+t \sum_{\sigma} (f^{\dagger}_{L \sigma} f_{R \sigma}+ {\rm h.c.} )
+\sum_{\alpha \sigma} \left[ v_{\alpha} \psi^{\dagger}_{\alpha \sigma}(0) f_{\alpha \sigma} + {\rm h.c.} \right],
\end{aligned}
\label{anderson}
\end{equation}
where $\epsilon$ is the dot energy and $U>0$ is the on-site interaction. $\alpha=L,R$ labels the dot or lead position, while $\sigma=\uparrow,\downarrow$ is the spin index. $t$ and $v_{\alpha}$ refer to the inter-dot tunneling strength and the hybridization between lead and dot, respectively.

When $\epsilon=-U/2$ and $U$ is much larger than any other energy scale, both dots are singly occupied. One can then project out the empty and doubly occupied states with a Schrieffer-Wolff transformation \cite{HewsonBooks}. In the second-order transformation in which two Anderson model tunneling terms are combined, one obtains Kondo and exchange tunneling processes given by Eq.\,(2) in the main text, with $J_{L/R}=4 (v_{L/R})^2/U$ and $K=4t^2/U$. When considering the third-order transformation, we will have the charge tunneling term with the tunneling strength given by $V_{L/R}=(16v_L v_R t)/U^2$ \citep{Logan2CKPRB11s}.

\section{II.\hspace{10pt}The critical Hamiltonian ($K \!=\! K_c$) in two dot Kondo
model}


The full Hamiltonian of the problem is given by Eq.\,(5) of the main 
text. However, at low energy scales and for $K \!=\! K_c$ it is possible to 
derive an effective model, Eq.\,(7) of the main text, that is analytically 
more convenient to work with. For convenience, we break $H \!=\! H_1 + H_2$ 
into two parts: 
(i) $H_1 \!=\! H^0_{\rm leads} + H_{\rm dots}$, and 
(ii) $H_2 \!=\! H_{LR} + H^0_{\varphi}$. 
We place $H^0_{\varphi}$ with $H_{LR}$ in $H_2$ because only $H_{LR}$ couples to dissipation. In this section we closely follow the treatment of the two-impurity Kondo model by Junwu Gan \cite{GanPRB95s}. 

\subsection{A.\hspace{8pt}Bosonization and unitary transformation}
\label{sec:bosonization}

The first step is to define symmetric and antisymmetric impurity
operators, $S^{\lambda}_+ \!=\! (S^{\lambda}_L + S^{\lambda}_R)$ and
$S^{\lambda}_- \!=\! (S^{\lambda}_L - S^{\lambda}_R)$, and rewrite
\begin{equation}
  H_1 = H_{leads}^0 + \sum_{\lambda} \frac{J}{2}  [(s^{\lambda}_L +
  s^{\lambda}_R) S^{\lambda}_+ + (s^{\lambda}_L - s^{\lambda}_R)
  S^{\lambda}_-] + K \vec{S}_L \cdot \vec{S}_R . \label{h1}
\end{equation}
We follow the standard bosonization prescription \cite{GogolinBooks}, using the Mandelstam identity,
\begin{equation}
  \psi_{\beta, \sigma} = \frac{F_{\beta, \sigma}}{\sqrt{2 \pi \alpha}} e^{i
  \phi_{\beta \sigma}}, \label{bosonization}
\end{equation}
where $\beta$ and $\sigma$ are lead and spin indices, respectively, and
$F_{\beta,\sigma}$ are Klein factors that preserve the fermionic 
anti-commutation relation for different fermionic flavors. The constant 
$\alpha$ is of order the inverse of the bare cut-off of the fermionic theory. 

It will be convenient to make spin-charge separation explicit by rotating the  bosonic fields,
\begin{equation}
\begin{aligned}
\phi_c &=(\phi_{L \uparrow}+\phi_{L \downarrow}+\phi_{R \uparrow}+\phi_{R \downarrow})/2, \qquad
&\phi_s =(\phi_{L \uparrow}-\phi_{L \downarrow}+\phi_{R \uparrow}-\phi_{R \downarrow})/2, \\
\phi_{cf} &=(\phi_{L \uparrow}+\phi_{L \downarrow}-\phi_{R \uparrow}-\phi_{R \downarrow})/2, \qquad
&\phi_{sf} =(\phi_{L, \uparrow}-\phi_{L \downarrow}-\phi_{R \uparrow}+\phi_{R \downarrow})/2.
\end{aligned}
\label{rotate}
\end{equation}
The charge field $\phi_c$ would couple to the fluctuation of the total charge on the two dots; since we consider the singly occupied regime of the quantum  dots in which there can be no charge fluctuations, this field decouples from the problem. The charge flavor field, $\phi_{cf}$, encodes charge transfer between the two leads, and thus is the bosonic field that couples to the environmental noise in our model. Similarly, $\phi_{s}$ and $\phi_{sf}$ correspond, respectively, to total spin and spin transfer between the leads.
Using these definitions, we find
\begin{equation}
\begin{aligned}
H_1=&\sum_{\beta=c,s,sf,cf} H^0_{\beta} 
+\frac{J}{2 \pi} \left\{ \partial_x \phi_s(0)S^z_++\partial_x \phi_{sf}(0)S^z_-  \right\}
+ K\, \vec{S}_L \cdot \vec{S}_R
\\
&+\frac{J}{\pi \alpha} \left\{ \cos\phi_{sf}(0) [\cos\phi_s(0)S^x_+ -\sin\phi_s(0)S^y_+]-\sin \phi_{sf}(0)[\sin\phi_s(0)S^x_-+\cos\phi_s(0)S^y_-]\right\}
\end{aligned}
\label{afterb}
\end{equation}
where $\sum_{\beta} H^0_{\beta}$ refers to the free bosonic Hamiltonians.

Because we are using Abelian bosonization, it is convenient to rotate the
system with the unitary $\hat{U} \!=\! e^{- iS^z_+ \phi_s(0)}$, a transformation that is well known in the Kondo literature. 
This transformation has several effects. First, its action on $S^x_{\pm}$ and $S^y_{\pm}$ leads to a decoupling of these operators from $\phi_s$. Second, its action on $\partial_x\phi_s(0)$ leads to two additional terms, 
$-\partial_x \phi_s(0) S^z_{+}$ and $-\frac{J-\pi}{\pi \alpha} (S^z_+)^2$, that change the coupling with $S^z_{L,R}$. The resulting form of the Hamiltonian, 
\begin{equation}
\begin{aligned}
&H_1=\sum_{\beta} H^0_{\beta} +\frac{J}{\pi \alpha} \left\{ \cos\phi_{sf}(0) S^x_+ -\sin \phi_{sf}(0)S^y_-\right\}+\frac{J-2 \pi}{2 \pi}  \partial_x \phi_s(0)S^z_+\\
&+\frac{J}{2 \pi}\partial_x \phi_{sf}(0)S^z_-
+ K\sum_{\lambda=x,y} S^{\lambda}_L S^{\lambda}_R
+ \Big[ K-\frac{2}{\pi \alpha} (J-\pi) \Big] S^z_L S^z_R , 
\end{aligned}
\label{afterb}
\end{equation}
is clearly highly anisotropic in the dot's spin degrees of freedom.

We now switch back to describing the leads using fermions, namely those defined by 
\begin{equation}
\begin{aligned}
&\psi_{sf}(x)=\frac{1}{\sqrt{2 \pi \alpha}} e^{i \phi_{sf}(x)},
\qquad \psi^{\dagger}_s(x)\psi_s(x)=\frac{1}{2 \pi} \partial_x \phi_s(x),
\\&\psi_{cf}(x) =\frac{1}{\sqrt{2\pi \alpha}}e^{i \phi_{cf}(x)}
e^{-i \pi \int_{-\infty}^{\infty}dx \psi^{\dagger}_{sf}(x)\psi_{sf}(x)},
\end{aligned}
\label{ref}
\end{equation}
which are not, of course, the original fermions because of the several rotations during the bosonic description. In terms of these new fermions,  $H_1$ is
\begin{equation}
\begin{aligned}
&H_1=\sum_{\lambda} K^{\lambda} S^{\lambda}_L S^{\lambda}_R+ (J-2 \pi) \psi^{\dagger}_s(0) \psi_s(0) S^z_++J \psi^{\dagger}_{sf}(0) \psi_{sf}(0)S^z_-
\\&+\frac{J}{2 \sqrt{2\pi \alpha}}\left\{ [\psi_{sf}(0)+\psi^{\dagger}_{sf}(0)]S^x_++i J[\psi_{sf}(0)-\psi^{\dagger}_{sf}(0)]S^y_- \right\},
\end{aligned}
\label{h1ref}
\end{equation}
where $K^x \!=\! K^y \!=\! K$ and $K^z \!=\! K - \frac{2}{\pi \alpha}  (J - \pi)$. 
Ref.\,\cite{GanPRB95s} points out that the second term in Eq.\,(\ref{h1ref}) 
only renormalizes the $K^z$ part of the original Hamiltonian, thus perhaps shifting the value of $K_c$ but not affecting the physics at $K \!=\! K_c$. We therefore drop this term from the Hamiltonian at this point.

\subsection{B.\hspace{8pt}Schrieffer-Wolff transformation}

\begin{figure}[b]
  \label{critical_Sup_Mat}
\hfill
\begin{minipage}[t]{3.5in}
  {\includegraphics[width=3.3in]{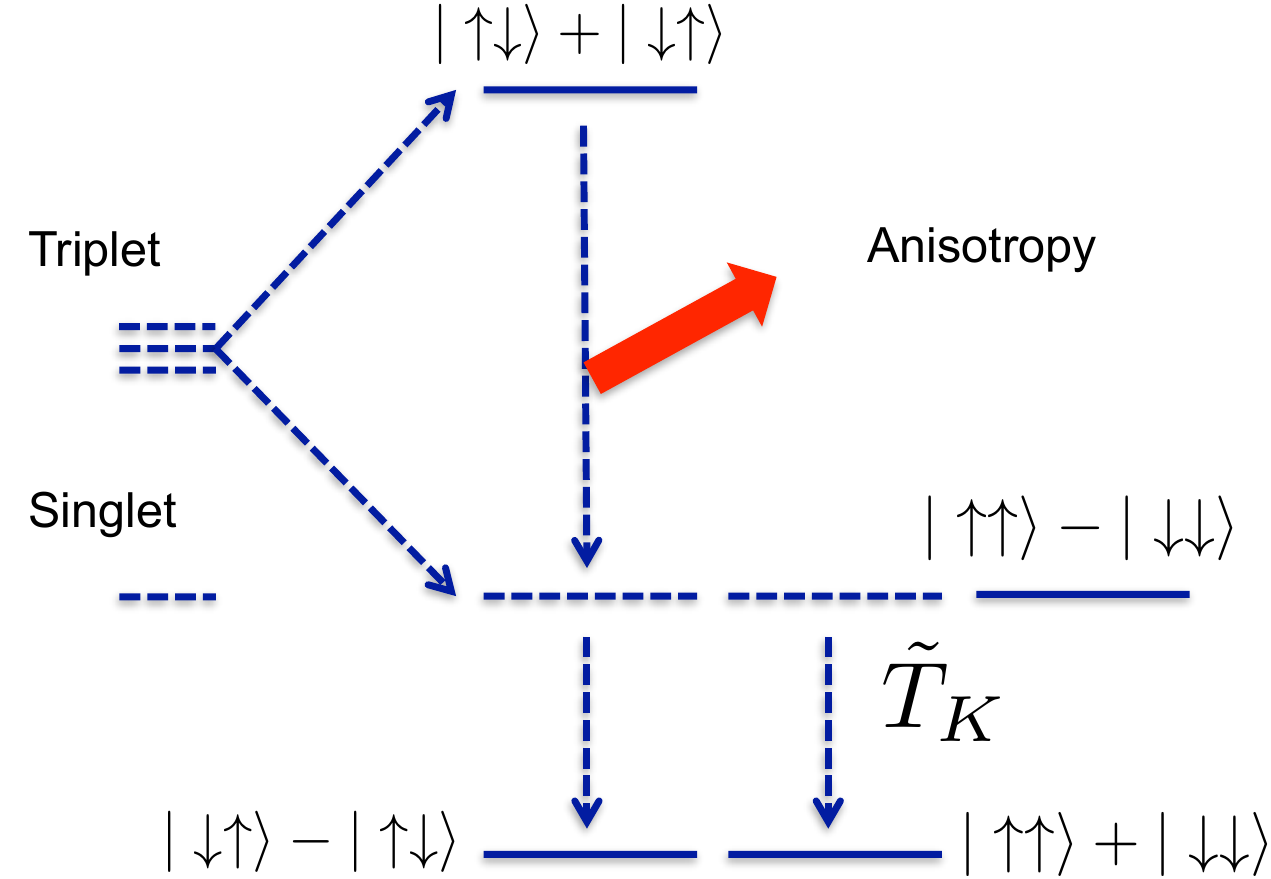}} 
\end{minipage}
\begin{minipage}[b]{2.9in}
  \caption{Low energy states of the system. One of the four states, 
  $|\! \uparrow
  \downarrow \rangle + |\! \uparrow \downarrow \rangle $,
  is lifted due to the anisotropy of $K^z$ and another state, $|\! \uparrow
  \uparrow \rangle - |\! \downarrow \downarrow \rangle $,
  has higher energy due to its decoupling in Eq.\,(\ref{h1ref}). The remaining doublet forms a two-dimensional low-energy space onto which we project via a Schrieffer-Wolff transformation.}
  \vspace*{0.8in}
\end{minipage}
\end{figure}

The main consequence of $\hat{U}$ is now evident \cite{GanPRB95s}. In this representation, the
symmetry in the exchange couplings is explicitly broken. These new local
degrees of freedom (dressed by $\hat{U}$) have a different energy level
structure than the original ones. There are four local states (see Fig.\,\ref{critical_Sup_Mat}) to consider: (i) The state 
$(|\!\!\uparrow\uparrow\rangle - |\!\!\downarrow\downarrow\rangle)/\sqrt{2}$ decouples from the fermionic leads and therefore its energy is not changed by the transformation. (ii) The triplet state 
$(|\!\downarrow\uparrow\rangle + |\!\uparrow\downarrow\rangle)/\sqrt{2}$ 
has its energy raised by exchange anisotropy. (iii) Finally, the
states 
$(|\!\uparrow\uparrow\rangle + |\!\downarrow\downarrow\rangle)/
\sqrt{2}$ and $(|\!\uparrow\downarrow\rangle - |\!\downarrow\uparrow
\rangle)/\sqrt{2}$ have their energy reduced. They are exactly degenerate at the critical coupling $K \!=\!K_c$; the value of $K_c$ for which this occurs depends, of course, on the value of $J$.

At the critical $K \!=\! K_c (J)$ Eq.\,(\ref{h1ref}) can be projected onto the
Hilbert space of these two lowest energy states through a Schrieffer-Wolff
transformation. First, a new set of local fermionic operators is defined,
\begin{equation}
\begin{aligned}
d^{\dagger}d|\Omega\rangle=0\ \to\ |\Omega\rangle=(|\uparrow\uparrow\rangle+|\downarrow\downarrow\rangle)/\sqrt{2},\quad 
d^{\dagger}d|\Omega'\rangle=|\Omega'\rangle\ \to\ |\Omega'\rangle=(|\uparrow\downarrow\rangle-|\downarrow\uparrow\rangle)/\sqrt{2}.
\end{aligned}
\label{dd}
\end{equation}
Then, the fermionic fields are rotated by $\tilde{\psi}_{\beta} (x) \!=\!
\psi_{\beta} (x) e^{- i \pi d^{\dagger} d}$ to ensure the anti-commutation
relations. 
Projecting onto this low-energy two-dimensional subspace yields
\begin{equation}
H_1=\sum_{\beta=\left\{ c,s,sf,cf,\varphi \right\}}H^0_{\beta}+\tilde{J}[\tilde{\psi}_{sf}(0)-\tilde{\psi}_{sf}^{\dagger}(0)] (d+d^{\dagger})-(K-K_c)d^{\dagger}d,
\label{hc}
\end{equation}
where 
\begin{equation}
\tilde{J} = \frac{J}{\sqrt{2\pi\alpha}} \left[ 1 + \frac{J}{4\pi\alpha(K_c + {\tilde T}_K)} \right].
\end{equation}
To validate the Schrieffer-Wolff transformation above, $K-K_c \ll \tilde{J}$ must be satisfied such that those two ground states are approximately degenerate.
$\tilde{T}_K$ here is the Kondo temperature that results from the leads screening the doublet described by $d$ and $d^{\dagger}$. The
$d^{\dagger} d$ term of Eq.\,(\ref{hc}) is a relevant perturbation (since it
breaks the degeneracy between the two states of Fig.\,\ref{critical_Sup_Mat}). Hence, for $K \!\neq\! K_c$ the system is driven away from the two-impurity Kondo fixed
point.

\subsection{C.\hspace{10pt}The charge transport operator}

All the steps that lead to Eq.\,(\ref{hc}) now must be applied to $H_2 \!=\!
H^0_{\varphi} + H_{LR}$. We start by disregarding the dissipative terms. Carrying out bosonization, unitary transformation, and refermionization as in section \ref{sec:bosonization}, we obtain
\begin{subequations}
\begin{eqnarray}
H_{LR} &=& V_{LR}\left\{ [\psi^{\dagger}_{cf}(0)+\psi_{cf}(0)] (-S_L^-S^z_R+S^z_L S^-_R-S^+_L S^z_R+S^z_LS^+_R)  \right. \label{h3fa}
\\ & &+ \; [\psi^{\dagger}_{cf}(0)\psi^{\dagger}_{sf}(0)-\psi^{\dagger}_{sf}(0)\psi_{cf}(0)] S^-_LS^+_R + {\rm h.c.} \label{h3fb}
\\& &+ \left. [\psi_{sf}(0)-\psi^{\dagger}_{sf}(0)][\psi_{cf}(0)+\psi^{\dagger}_{cf}(0)] S^z_L S^z_R \right\}.
\label{h3fc}
\end{eqnarray}
\end{subequations}
To carry out the Schrieffer-Wolff transformation, we first note how each of the local operators in this expression acts on the low energy doublet of local degrees of freedom: 
\begin{equation}
\begin{aligned}
&(-S_L^-S^z_R+S^z_L S^-_R-S^{+}_LS^z_R+S^z_LS^+_R) (|\uparrow\uparrow\rangle+|\downarrow\downarrow\rangle)/\sqrt{2}=(|\uparrow\downarrow\rangle-|\downarrow\uparrow\rangle)/\sqrt{2},
\\&(-S_L^-S^z_R+S^z_L S^-_R-S^{+}_LS^z_R+S^z_LS^+_R) (|\uparrow\downarrow\rangle-|\downarrow\uparrow\rangle)/\sqrt{2}=(|\uparrow\uparrow\rangle+|\downarrow\downarrow\rangle)/\sqrt{2},
\end{aligned}
\label{swt1}
\end{equation}

\begin{equation}
\begin{aligned}
&(S_L^-S_R^+) (|\uparrow\uparrow\rangle+|\downarrow\downarrow\rangle)/\sqrt{2}=0,
\\&(S_L^-S_R^+) (|\uparrow\downarrow\rangle-|\downarrow\uparrow\rangle)/\sqrt{2}=|\downarrow\uparrow\rangle/\sqrt{2},
\end{aligned}
\label{swt2}
\end{equation}

\begin{equation}
\begin{aligned}
&(S_L^zS_R^z) (|\uparrow\uparrow\rangle+|\downarrow\downarrow\rangle)/\sqrt{2}= \frac{(|\uparrow\uparrow\rangle+|\downarrow\downarrow\rangle)}{4\sqrt{2}},
\\&(S_L^zS_R^z) (|\uparrow\downarrow\rangle-|\downarrow\uparrow\rangle)/\sqrt{2}=\frac{(|\uparrow\downarrow\rangle-|\downarrow\uparrow\rangle)}{4 \sqrt{2}},
\end{aligned}
\label{swt3}
\end{equation}
Eq.\,(\ref{swt1}) shows that the local operator appearing in Eq.\,(\ref{h3fa}),
$- S_L^- S^z_R + S^z_L S^-_R - S^{+}_L S^z_R
+ S^z_L S^+_R$, acts on the doublet as $(d + d^{\dagger})$. 
Turning to Eq.\,(\ref{h3fb}), we see from (\ref{swt2}) that 
$S_L^- S_R^+$ is proportional to $d^{\dagger} d$. Notice that the lead operators that couple to it have scaling dimension one, and we will see that $d^{\dagger} d$ has scaling dimension 1/2 around the critical point [see the first paragraph after Eq.\,(7) of the main text]. Thus, the first order projection of $S_L^- S_R^+$ is RG irrelevant, and we need consider only
higher order projections. One can check that
\begin{equation}
\begin{aligned}
\hat{Q}S^z_-(1-\hat{Q})S^-_L S^+_R (1-\hat{Q})S^x_+ \hat{Q}=\frac{1}{2}d^{\dagger}=\hat{Q}S^z_-(1-\hat{Q})S^-_R S^+_L (1-\hat{Q})S^x_+ \hat{Q},
\\ \hat{Q}S^x_+(1-\hat{Q})S^-_L S^+_R (1-\hat{Q})S^z_- \hat{Q}=\frac{1}{2}d=\hat{Q}S^x_+(1-\hat{Q})S^-_R S^+_L (1-\hat{Q})S^z_- \hat{Q},
\end{aligned}
\label{swt2s}
\end{equation}
where $\hat{Q}$ is the projection operator onto the local doublet. In calculating the associated lead operator, one finds cancellations such that
(\ref{h3fb}) corresponds to an operator
proportional to $(d + d^{\dagger})  [\psi^{\dagger}_{cf} (0) + \psi_{cf}
(0)]$. Finally, the transformation of (\ref{h3fc}) 
is similar to that of (\ref{h3fb}), yielding a term in first-order of form
$(d^{\dagger} d - 1 / 2)$ and in second-order again 
$(d + d^{\dagger})  [\psi^{\dagger}_{cf} (0) + \psi_{cf} (0)]$.

The last step in this derivation is to recall that $\tilde{\psi}_{\beta} (x)
\!=\! \psi_{\beta} (x) e^{- i \pi d^{\dagger} d}$; this changes the
operator $(d + d^{\dagger})$ to $(d - d^{\dagger})$. Thus finally, the
effective low-energy Hamiltonian for charge transport between the leads is
\begin{equation}
  \tilde{H}_{LR} = \tilde{V}_{LR}  [ \tilde{\psi}^{\dagger}_{cf} (0) +
  \tilde{\psi}_{cf} (0)]  (d - d^{\dagger}), 
  \label{effcharge}
\end{equation}
where 
\begin{equation}
\tilde{V}_{LR}= V_{LR} \left[ 1 + \frac{J^2}{8 (K_c+\tilde{T}_k)^2 (2 \pi \alpha)^{5/2}} \right]
\end{equation} 
is the renormalized tunneling strength. As a reminder, if we start with the simplified tunneling Hamiltonian $\hat{H}_{LR}$ as mentioned in the main text, we will end with an effective Hamiltonian that shares the form of Eq.\,(\ref{effcharge}) but with different $\tilde{V}_{LR}$. That is why we claim that the simplified $\hat{H}_{LR}$ catches the physics around the $K=K_c$ critical point.

In order to introduce dissipation (in the following section), it is convenient to express the Hamiltonian in bosonic variables, so we use the bosonization identity 
Eq.\,(\ref{bosonization}) again. The bosonic fields thus introduced are not the same as those of Section \ref{sec:bosonization} because of the Jordan-Wigner string operator used in obtaining Eq.\,(\ref{effcharge}). Nevertheless, we use the same notation for these fields in order to avoid complicating the notation. 
Combining all the terms, we see that the Hamiltonian is 
\begin{equation}
\begin{aligned}
H =&\sum_{\beta=\left\{ c,s,cf,sf,\varphi \right\}}\!\!\! H^0_{\beta}
+2i\tilde{J} \frac{F_{sf}}{\sqrt{2\pi \alpha}}\sin\left[ \phi_{sf}(0) \right] (d+d^{\dagger})+2\tilde{V}_{LR} \frac{F_{cf}}{\sqrt{2\pi \alpha}}\cos[\phi_{cf}(0)](d-d^{\dagger})\\
&\qquad\qquad -(K-K_c)d^{\dagger}d.
\end{aligned}
\label{hc0diss}
\end{equation}
As in the case of the $K - K_c$ term, the $\tilde{V}_{LR}$ term is a relevant operator
that drives the system away from the two-impurity Kondo fixed point. As mentioned in previous text, $K-K_c$ is a small quantity such that two ground states are approximately degenerate.

\subsection{D.\hspace{8pt}Introducing dissipation}

The environmental noise enters through the field $\varphi$. The charge shift operator $e^{i\sqrt{2r}\varphi(0)}$ naturally couples with the lead field representing charge transfer, namely $\phi_{cf}$. Since $\varphi$ is unaffected by the transformations of the previous section, it can be introduced by replacing $\phi_{cf}$ by $\phi_{cf} + \sqrt{2r} \varphi$ in 
the tunneling Hamiltonian. When $r \neq 0$, we cannot apply the refermionization procedure and are thus forced to work with the bosonic Hamiltonian,
\begin{equation}
\begin{aligned}
H =&\sum_{\beta=\left\{ c,s,cf,sf,\varphi \right\}}\!\!\!\!\!\! H^0_{\beta}
+2i\tilde{J}\frac{F_{sf}}{\sqrt{2\pi \alpha}}\sin\left[ \phi_{sf}(0) \right] (d+d^{\dagger})+2\tilde{V}_{LR}\frac{F_{cf}}{\sqrt{2\pi\alpha}} \cos[\phi_{cf}(0)+\sqrt{2r}\varphi(0)](d-d^{\dagger})\\
&-(K-K_c)d^{\dagger}d.
\end{aligned}
\label{hcdiss}
\end{equation}
For $K \!=\! K_c$, this is Eq.\,(7) of the main text.

\section{III.\hspace{10pt}Conductance at the Intermediate Fixed Points}

\subsection{A.\hspace{8pt}Conductance at zero temperature}

Conductance is naturally related to scattering states via, e.g., the Landauer-B\"uttiker approach \cite{NazarovBooks}. Following Refs.\,\cite{SelaPairTunnelPRL09as,MaleckiDoubdotPRB10s,MaleckiErratumPRB11s}, we use here the scattering states 
\begin{equation}
  \Psi_{j \sigma} (x) = \theta(x) \psi_{j \sigma} (x) + \sum_{j'} \theta (- x) s_{jj'}
  \psi_{j' \sigma} (x), \label{scatts}
\end{equation}
where $\theta (x)$ refers to an incoming fermionic state and $\theta (- x)$
refers to the scattered state. The $j$ and $j'$ subscripts label the leads and
allow us to write the scattering matrix
\begin{equation}
  s = \left( \begin{array}{cc}
    \cos 2 \delta & - i \sin 2 \delta\\
    - i \sin 2 \delta & \cos 2 \delta
  \end{array} \right),
\end{equation}
where $\delta \!=\! \frac{1}{2} 
\arg \Big( \sqrt{T_{\delta K}/T_K} + i\sqrt{T_{\rm noise}^{LR}/T_K}
\;\Big)$
is the scattering phase shift \cite{SelaPairTunnelPRL09as,MaleckiDoubdotPRB10s,MaleckiErratumPRB11s}. The crossover temperatures $T_{\delta K}$ and $T_{\rm noise}^{LR}$ are defined in the main text, and $T_K$ is the Kondo temperature associated with the bare dot-lead coupling. 


At \IFPtwo, the phase shift is $\delta \!=\! \pi / 4$. This has a strong
consequence on the process that transfers two electrons simultaneously between the two leads, $\Psi^{\dagger}_{L \uparrow}
\Psi^{\dagger}_{L \downarrow} \Psi_{R \uparrow} \Psi_{R \downarrow}$ \cite{SelaPairTunnelPRL09as,MaleckiDoubdotPRB10s,MaleckiErratumPRB11s}. 
Using Eq.\,(\ref{scatts}), one finds that this process can be written as
\begin{equation}
  \Psi^{\dagger}_{L, \uparrow} \Psi^{\dagger}_{L, \downarrow} \Psi_{R,
  \uparrow} \Psi_{R, \downarrow} (x) \theta (x) + \Psi^{\dagger}_{L, \uparrow}
  \Psi^{\dagger}_{L, \downarrow} \Psi_{R, \uparrow} \Psi_{R, \downarrow} (x)
  \theta (- x) . \label{2e}
\end{equation}
This shows that the ``incoming'' state coincides exactly with the ``outgoing'' state. This perfect transmission implies that there is zero shot noise during this scattering process \cite{NazarovBooks}, which in turn implies that charge transport processes will not be affected by dissipation \cite{LevyYeyatiPRL01s,SafiSaleurPRL04s}. Thus, we expect perfect conductance at zero temperature.

\subsection{B.\hspace{8pt}Low temperature conductance}

Refs.\,\citep{GogolinBooks,KaneFisherPRB92s} show that the behavior of the linear conductance can be
calculated by perturbation theory improved by RG. At low temperatures,
the result is $G \propto T^{2(\chi - 1)}$, where $\chi$
is the scaling dimension of the leading irrelevant operator.

For $r > 1 / 2$, \IFPone\ is the intermediate fixed point. The leading irrelevant operator has scaling dimension $(1+2r)/2$. Thus, near \IFPone\ the
temperature dependence of the conductance is $G \propto T^{2 r - 1}$.

For $r < 1 / 2$, \IFPtwo\ is the intermediate fixed point. The leading irrelevant
operator is obtained by a duality transformation on the charge transport
operator \citep{GogolinBooks,KaneFisherPRB92s}. The scaling dimension of the dual operator
is $2/(1 + 2 r)$ , and therefore the conductance is 
$1 - G \propto T^{2 \frac{1 - 2 r}{1 + 2 r}}$ .

\section{IV.\hspace{10pt}Boundary contribution to the zero temperature
entropy}

The boundary sine-Gordon model is given by the Hamiltonian
\begin{equation}
H = \int_{0}^{\infty}\frac{dx}{8\pi}
\left\{ (\partial_{x}\Phi)^{2}+\Pi^{2}(x) \right\}
+ \lambda\cos\left[\frac{\sqrt{g^*}}{2}\Phi(0)\right],
\end{equation}
where $g^*$ is known as the compactification ratio and the field $\Phi$
and $\Pi$ are periodic variables. The ground state degeneracy, 
$g(\Phi)$, of this model was evaluated 
in Refs.\,\cite{WongAffleck94s,OshikawaAffleck97s}
for the two conformally invariant boundary conditions,
\begin{eqnarray}
g_{D}(\Phi) & = & \frac{1}{\sqrt[4]{g^*}},\\
g_{N}(\Phi) & = & \frac{\sqrt[4]{g^*}}{\sqrt{2}},
\end{eqnarray}
where $D$ stands for Dirichlet and $N$ for Neumann. The model can
be rewritten with a single chiral bosonic field by analytic continuation. 

In the main text, the chiral bosonic fields $\tilde{\varphi}$, $\phi_{c}$,
$\phi_{s}$ and $\phi_{sf}$ have $g^*\!=\!1$, whereas $\phi_{cf}$ has
$g^*\!=\!1+2r$. The local degrees of freedom also contribute to the ground
state degeneracy with $g_{\rm dots}\!=\!2$ 
\footnote{Ref.\,\cite{GogolinBooks} p. 397}.

We can now apply these known results to the two main fixed points
discussed in the manuscript. 
\begin{itemize}
\item \IFPone: $\phi_{sf}$ has a Neumann boundary condition, while
$\tilde{\varphi}$, $\phi_{c}$, $\phi_{s}$, and $\tilde\phi_{cf}$ have 
Dirichlet boundary conditions.  Hence, the total ground state degeneracy is 
\begin{eqnarray}
g_{{\rm IFP}_1} & = & g(\tilde{\varphi})g(\phi_{c})g(\phi_{s})g(\tilde\phi_{cf})g(\phi_{sf})g_{\rm dots}\nonumber \\
 & = & \sqrt[4]{\frac{4}{1+2r}}\label{eq:entropyIFP1} \;.
\end{eqnarray}
 
\item \IFPtwo: $\phi_{sf}$ and $\tilde\phi_{cf}$ have Neumann boundary condition,
while $\tilde{\varphi}$, $\phi_{c}$, and $\phi_{s}$ have Dirichlet
boundary condition. The total ground state degeneracy for this case
is 
\end{itemize}
\begin{eqnarray}
g_{{\rm IFP}_2} & = & \sqrt[4]{1+2r} \;.
\label{eq:entropyIFP2}
\end{eqnarray}

{\def\bibfont{\normalsize}
\bibliography{supp}}

\begin{thebibliography}{63}
\bibitem [{TKa()}]{TKappears}%
  \BibitemOpen
  \href@noop {} {}\bibinfo {note} {$T_K$ here is the Kondo temperature of the
  original problem, Eq.\,(\ref{eq:Hdots}). It appears as the high-energy cutoff
  for the scaling of $\tilde V_{LR}$: a Kondo state must form to have
  competition between the KFP and LSFP. For $r\!=\!0$, $T_\text{noise}^{LR}$ is
  $T^*$ introduced above.}\BibitemShut {Stop}%
\end{thebibliography}

\end{document}